\documentclass[a4paper]{jpconf}
\usepackage{graphicx}
\begin{document}
\title{Measurements of non-photonic electrons with the STAR experiment}

\author{Olga Rus\v n\' akov\' a for the STAR Collaboration}

\address{Czech Technical University in Prague\\ Faculty of Nuclear Sciences and Physical Engineering\\ B\v rehov\'a 7, 11519, Prague 1, Czech Republic}

\ead{olga.hajkova@fjfi.cvut.cz}

\begin{abstract}
Non-photonic electrons (NPE), produced by semileptonic decays of D and B mesons, are good probes to study the properties of hot and dense medium created in relativistic heavy ion collisions at RHIC. Studies of heavy quark production in p+p collisions can test the validity of perturbative QCD. They also provide a baseline to study the effects of nuclear matter on the production of heavy quarks in heavy ion collisions. In this paper, we present recent results of NPE spectra measured in p+p collisions at $\sqrt{s}=200$ GeV in mid-rapidity. We also report NPE nuclear modification factor $R_{AA}$ and elliptic flow $v_2$ in Au+Au collisions at $\sqrt{s_{NN}}=200$ GeV.

\end{abstract}

\section{Introduction}
Heavy quarks are good probes to study the properties of strongly interacting matter, Quark-Gluon Plasma (QGP). Due to their large masses, they are produced mainly during initial hard parton scatterings at RHIC, before the QGP phase. They are expected to interact with the medium differently from the light quarks. Hot nuclear matter effects, such as heavy quark inteactions with the QGP, change the heavy quark kinematic distribution but do not change the total heavy quark yield. Due to energy loss of heavy quarks in the QGP, their transverse momentum $p_{T}$ distributions in heavy ion collisions may fall steeper than those in p+p collisions. At small $p_{T}$, heavy quarks may thermalize with the medium and exhibit collective flow effects \cite{teory}.

At RHIC, heavy quarks could be studied by measuring NPE which are produced from semileptonic decays of D and B mesons. Measurements of NPE provide information on heavy quarks energy loss and elliptic flow in the hot and dense nuclear matter created in relativistic heavy ion collisions. 

Hot nuclear matter effects could be quantified by the nuclear modification factor ($R_{AA}$). $R_{AA}$ is defined as the ratio of the number of particles produced in nucleus-nucleus collisions to that in proton-proton collisions, scaled by the average number of binary collisions for a given centrality. $R_{AA} < 1$ may indicate heavy quark energy loss. In the case there are no medium effects the nuclear modification factor will be equal to unity.

\section{Analysis}
Data reported in this proceedings were collected in p+p collisions at $\sqrt{s}=200$ GeV in 2009, and in Au+Au collisions at $\sqrt{s_{NN}}=200$ GeV in 2010 with Minimum Bias (MB) and High Tower (HT) triggers, where MB triggered data are used for $p_T<2$ GeV/$c$ results and High Tower triggered data for $p_T>2$ GeV$/$c results. 

The Time Projection Chamber (TPC) is the primary tracking device for charged particle momentum determination at STAR. Information from the TPC was used for particle identification together with information from the Barrel Electromagnetic Calorimeter (BEMC), and the Time Of Flight (TOF) detector. The BEMC was used for high $p_T$ electron energy measurement and online trigger. The TOF detector was used to reduce hadron contamination at low $p_T$. 

Electron candidates are identified via ionization energy loss measured by the TPC combined with TOF velocity information, which together provide good PID capability at low-$p_T$ region. Electrons at high-$p_T$ are selected using the ratio of the track momentum and the energy deposited in the BEMC, the BSMD shower profile, and the distance between TPC track projected position at BEMC and reconstructed BEMC cluster position. The raw NPE yield is calculated as: $N_{NPE} = N_{Inc}*\epsilon_{pur} - N_{PHE}/\epsilon_{PHE}$, where $N_{NPE}$ is NPE yield, $N_{Inc}$ represents inclusive electron candidate yield, $\epsilon_{pur}$ is the purity of inclusive electron sample,  $N_{PHE}$ is the yield of reconstructed photonic electron background, which mainly originates from photon conversion in the detector material and from Dalitz decay of $\pi^0$ or $\eta$ mesons, and $\epsilon_{PHE}$ represents the photonic electron reconstruction efficiency. $\epsilon_{PHE}$ is estimated by full GEANT simulation. Finally, NPE yield is corrected by detector acceptance and efficiencies.

\section{NPE results in p+p collisions at $\sqrt{s}=200$ GeV}
\begin{figure}[h]
\begin{minipage}{16pc}
\includegraphics[width=19pc]{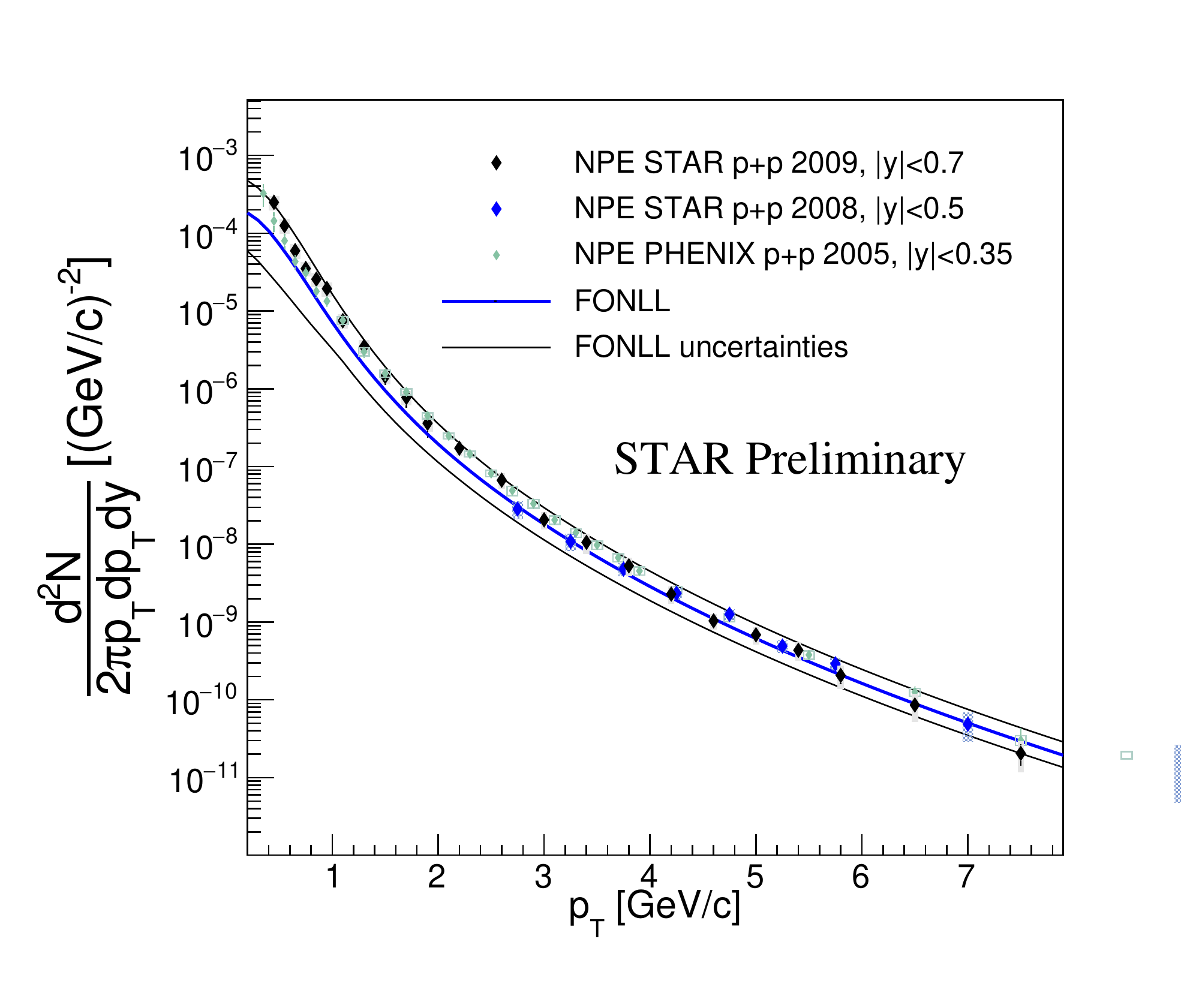}
\end{minipage}\hspace{2pc}%
\begin{minipage}{16pc}
\includegraphics[width=19pc]{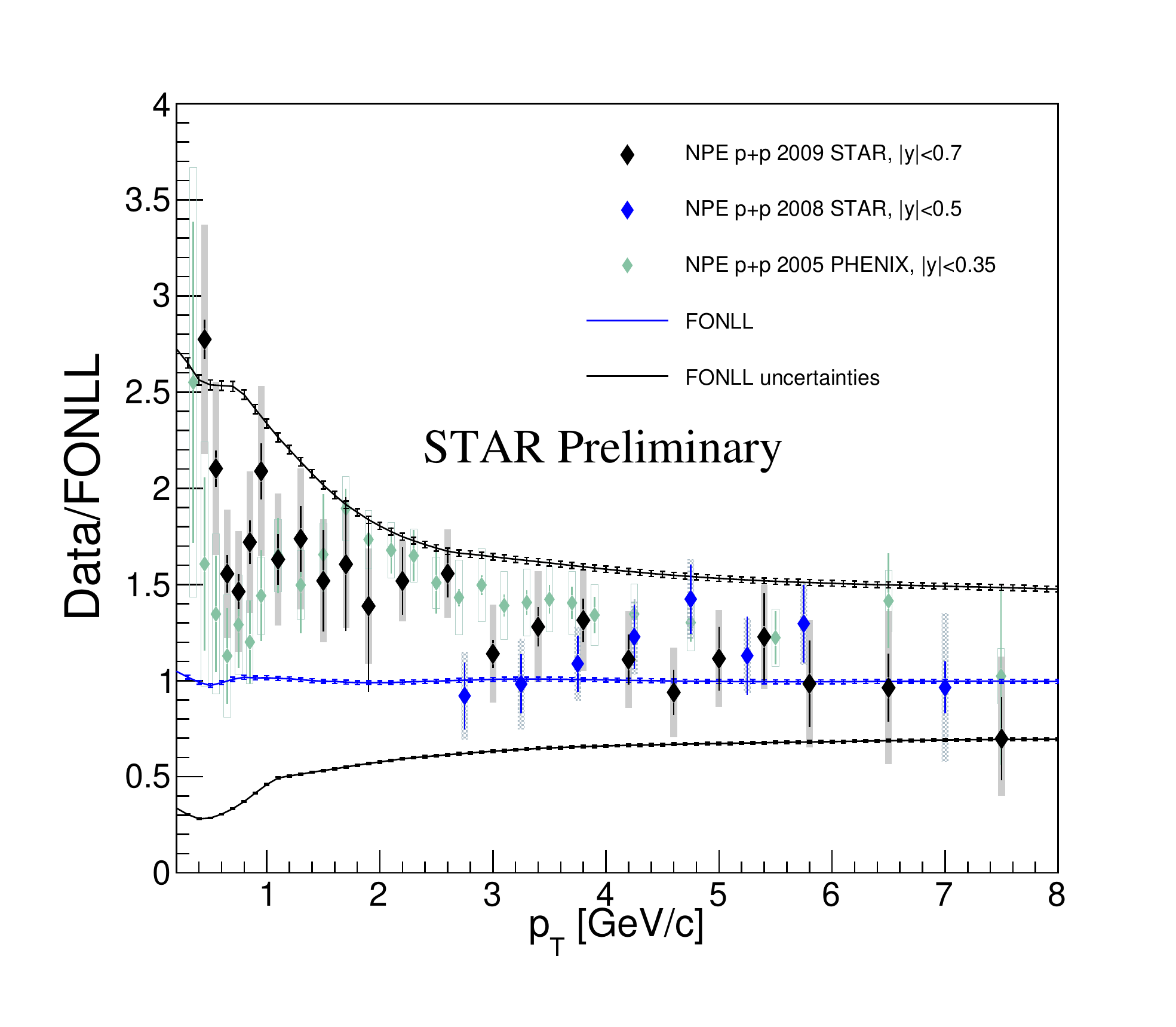}
\end{minipage} 
\caption{\label{label}Non-photonic electron yield (left) and ratio of data over FONLL calculation (right) \cite{FONLL}, STAR results from p+p collisions at $\sqrt{s}$=200 GeV from the year 2009 (black) compared with STAR data from the year 2008 (blue) \cite{clanek_pp} and PHENIX p+p results \cite{clanek_PHENIX}.}
\end{figure}

The recent results of STAR NPE measurement in p+p colisions at $\sqrt{s}=200$ GeV are shown in Fig. \ref{label}. Left panel presents the measured NPE yield as a function of $p_T$, where black marks show results obtained in the 2009 analysis and blue points show former analysis results \cite{clanek_pp}. The new data extends to lower $p_T$ region compared to the previous STAR measurement. Green points represent data from PHENIX (pseudorapidity range $|\eta|<0.35$) \cite{clanek_PHENIX}. STAR and PHENIX results are consistent in the overlappping $p_T$ range. STAR results are also compared with pQCD FONLL calculation (Fixed Order plus Next-to-Leading Logarithms \cite{FONLL}), where the FONLL central result is presented by the blue solid line and its upper and lower uncertainties by black lines. Experimental results are in good agreement with FONLL calculation and lie between its central value and the upper limit. 

Data and FONLL calculation are also compared in Fig. \ref{label} (right panel) where a ratio to FONLL central values is plotted.

\section{NPE results in Au+Au collisions at $\sqrt{s_{NN}}=200$ GeV}
\begin{figure}[h]
\begin{minipage}{16pc}
\includegraphics[width=17pc]{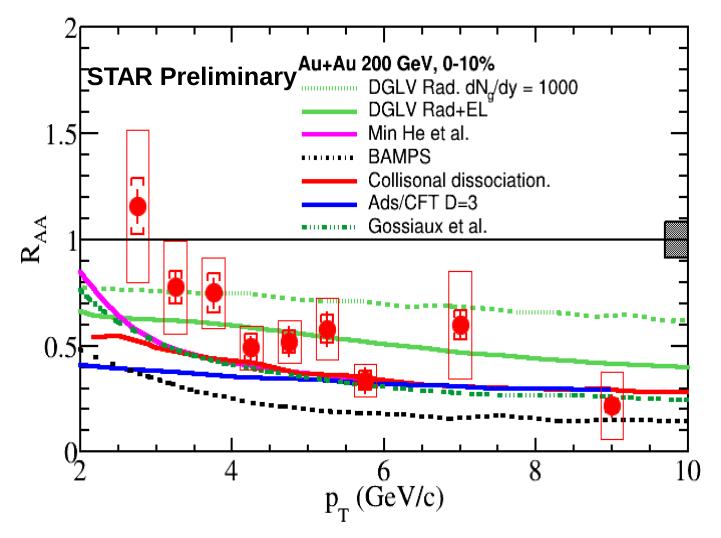}
\end{minipage}\hspace{2pc}%
\begin{minipage}{16pc}
\includegraphics[width=18pc]{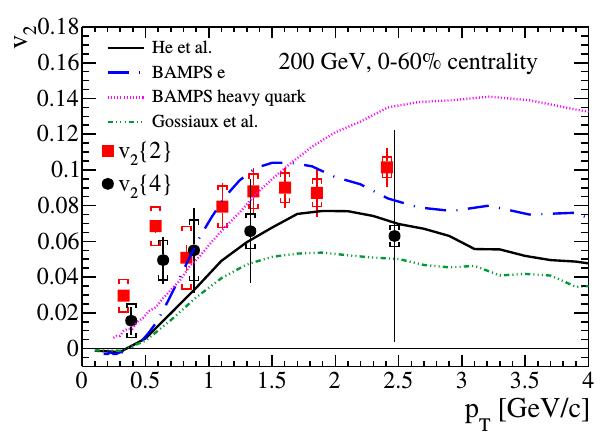}
\end{minipage} 
\caption{\label{label2}Non-photonic electron results from Au+Au collisions at $\sqrt{s}$=200 GeV from 2010. On the left plot is the NPE nuclear modification factor and on the right plot is the NPE elliptic flow. Both results are compared with theoretical models \cite{raa_1}-\cite{gos}. }
\end{figure}

In this section the recent results of NPE measurements in Au+Au collisions at $\sqrt{s_{NN}}$=200 GeV will be discussed. The NPE invariant yields were compared with FONLL calculations scaled by the average number of binary collisions in five centrality bins. In central and semicentral Au+Au collisions we observed a suppresion of NPE production compared to the FONLL calculation \cite{FONLL}.  
 Figure \ref{label2} (left) shows NPE $R_{AA}$ for the 0-10\% most central collisions compared to several theoretical models of heavy quark energy loss \cite{raa_1}-\cite{raa_5}. Gluon radiation scenario alone \cite{raa_1} (dashed green line) fails to explain the large NPE suppression at high $p_T$. When the collisional energy loss is added model calculations describe the data better. The collisional dissociation model \cite{raa_4} (red line) and the AdS/CFT calculation \cite{raa_5} (blue line) also describe the data well. Note that the baseline for nuclear modification factor calculation is produced from a combination of NPE spectra measured in the years 2005 and 2008 \cite{clanek_pp}. By including the aforementioned new p+p results from 2009 data, the $R_{AA}$ uncertainties will be reduced.
 
The right panel of Fig. \ref{label2} shows a measurements of NPE $v_2$ in Au+Au collisions at $\sqrt{s_{NN}}$=200 GeV. These results are obtained using 2-particle ($v_2\{2\}$) and 4-particle ($v_2\{4\}$) correlations \cite{flow}. These results are compared with theoretical models \cite{v2_1}-\cite{gos}. We observe finite $v_2$ at low $p_T$ and at high $p_T$ we observe increasing of $v_2$ which can arise from non-flow effects such as jet-like correlations.

\section{Summary}
The NPE invariant yield in p+p collisions at $\sqrt{s}=200$ GeV has been measured by the STAR experiment using high statistics data. The result can be described by FONLL calculations. Compared to the previously published STAR results, the new measurement using 2009 data extends the NPE spectrum to lower $p_T$ region. Results of NPE measurements in Au+Au collisions at $\sqrt{s_{NN}}$=200 GeV from STAR show large suppression of NPE production in central Au+Au collisions which cannnot be explained by gluon radiation energy loss alone. Finite NPE $v_2$ is observed at low $p_T$ which together with strong NPE suppresion in central collisions indicates strong charm-medium interaction.

\section{Outlook}
\begin{figure}[h]
\includegraphics[width=18pc]{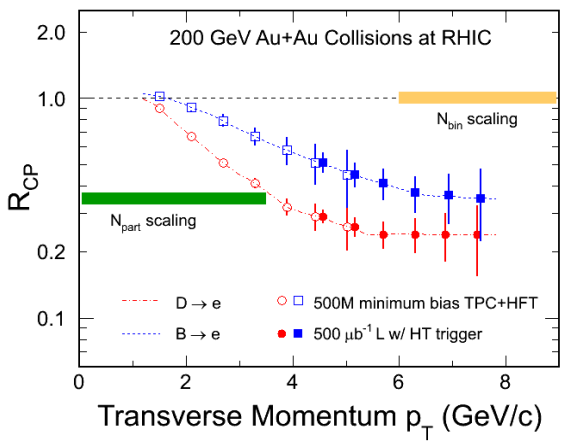}\hspace{2pc}%
\begin{minipage}[b]{14pc}\caption{\label{label3}Projection of $R_{CP}$ of non-photonic electrons from D (red) and B (blue) mesons separately \cite{HFT}. Open symbols are for 500M Au+Au simulated minimum-bias events, and filled symbols are for HT data with 500 $\mu$b$^{-1}$ sampled luminosity.}
\end{minipage}
\end{figure}

The new STAR detector, Heavy Flavor Tracker (HFT), which started its operation in 2014, extends the STAR particle identification capability to heavy flavor particles (i.e. particles containing heavy quarks). The HFT is able to topologically reconstruct charm mesons and baryons. This is made possible by reconstruction of secondary vertices with high precision. For NPE measurements it will be possible to distinguish between electrons from D and B mesons. Such studies will improve our understanding of heavy quarks interaction with QGP. Projected $R_{CP}$ results for electrons from these two sources are shown in Fig. \ref{label3}, where $R_{CP}$ is a central to peripheral nuclear modification factor, define as a ratio of yield in central collisions to yield in peripheral collisions scaled by number of binary collisions.
 
\section*{Acknowledgments}
This work was supported by Grant Agency of the Czech Technical University in Prague, grant No. SGS13/215/OHK4/3T/14 and by Grant Agency of the Czech Republic, grant No.13-20841S.

\end{document}